\documentclass[twocolumn,aps,showpacs,preprintnumbers,fleqn,superscriptaddress]{revtex4}
\usepackage{epsfig}
\usepackage{graphicx}
\usepackage{amsfonts}
\usepackage{subfigure}
\usepackage[dvips]{color}
\newcommand{\ig}{\includegraphics}
\newcommand{\ct}{\cite}
\newcommand{\bi}{\bibitem}

\newcommand{\ket}{\rangle}
\newcommand{\non}{\nonumber}

\newcommand{\be}{\begin{equation}}
\newcommand{\ee}{\end{equation}}
\newcommand{\ba}{\begin{eqnarray}}
\newcommand{\ea}{\end{eqnarray}}

\begin{document}

\title{Adiabatic dynamics in passage across quantum critical lines and gapless phases}
\author{Debanjan Chowdhury}
\email{debanjanchowdhury@gmail.com}
\author{Uma Divakaran}
\email{udiva@iitk.ac.in}
\author{Amit Dutta}
\email{dutta@iitk.ac.in}
\affiliation{Department of Physics, Indian Institute of Technology, Kanpur 208 016, India}

\date{\today}
\begin{abstract}
It is well known that the dynamics of a  quantum system is always non-adiabatic in passage
through a quantum critical point and the defect density in the final state following a quench
shows a power-law scaling  with the rate of quenching.
However, we propose here 
a possible situation where the  dynamics of a quantum system in passage across  
quantum critical regions is adiabatic and the defect density decays exponentially.  
This is achieved by incorporating additional interactions which lead to quantum
critical behavior and gapless phases but do not participate in the time evolution of the
system. To illustrate the general argument, we study the defect generation in the quantum critical 
dynamics of a spin-1/2 anisotropic  quantum XY spin chain with three spin interactions
and a linearly driven staggered magnetic field. 

\end{abstract}

\maketitle

Understanding the dynamics of a quantum system passing through a quantum
critical point \ct{sachdev99,dutta96}  has been a very active and fascinating area of research in recent 
years \ct{sengupta04,zurek05,damski06,polkovnikov05,polkovnikov07,
levitov06,mukherjee07,divakaran09,
sengupta08,dziarmaga06,dziarmaga05,
patane08}. 
The dynamical evolution can be initiated either by a 
sudden change  of a parameter in the 
Hamiltonian, called a sudden quench \ct{sengupta04}, or by 
a slow (adiabatic) quenching of a parameter\ct{zurek05,polkovnikov05}. The related entanglement
and fidelity properties are also being looked at \ct{fidelity}. It is well known  that
when  a quantum system, initially prepared in its ground state,  is swept adiabatically across a quantum critical point,
defects are generated in the final state of the system due to the critical slowing down which forces the dynamics of the system
to be non-adiabatic in the vicinity of the quantum critical point. In a linear passage through an isolated critical point, when a parameter ($e.g.,$ the magnetic field) is changed in time $t$ as $t/\tau$,
 the defect density ($n$) in the final state scales with the quenching
rate $1/\tau$ following the Kibble-Zurek scaling relation  \ct{kibble76,zurek96} given
by $n \sim \tau^{-d \nu /(\nu z+1)}$ in the adiabatic limit ($\tau\to \infty$)\ct{zurek05,polkovnikov05,polkovnikov07}.
Here,  $d$ is the spatial dimension, $\nu$ and $z$ are the correlation length and
the dynamical exponents, respectively, associated with the quantum critical point \ct{sachdev99}. 
The experimental verification of such dynamics is now possible by studying the dynamics of
 atoms trapped in optical lattices \ct{bloch07}.

In this work, we explore the possibility of adiabatic dynamics or exponentially decaying defect
density even in passage through quantum critical points and gapless phases.
We show that this occurs only in a special situation in which an additional term of  the 
Hamiltonian that leads to the quantum critical behavior and gapless phases  does not 
participate in the  dynamics of the system.  We illustrate the 
 general argument using  a quantum spin-1/2 XY chain with multispin interactions and a 
staggered magnetic field which is exactly solvable via the Jordan-Wigner transformation.
We thus provide an example of a special situation that contradicts the power-law scaling of the defect density  for a non-random quantum system \ct{zurek05,polkovnikov05}.

Let us begin the discussion with a  $d$-dimensional quantum Hamiltonian given by

\ba
H &=& \sum_k \psi^{\dagger}(\vec k) H(\vec k) \psi(\vec k); \nonumber\\
H(\vec k)&=& \alpha c(\vec k) \hat {1} + \left(\lambda \sigma^z+ \Delta (\vec k) \sigma_{+} + \Delta^{*}(\vec k) \sigma_{-} \right)
\label{ham1}
\ea 
where $\alpha$, $c(\vec k)$ and $\Delta (\vec k)$ are model dependent functions, $\sigma^{i}(i=x,y,z)$ are Pauli spin matrices,
 $\sigma_{+} = (\sigma^x + i \sigma^y)/2$, 
$\sigma_{-} = (\sigma^x - i \sigma^y)/2$ and $\hat 1$ denotes 
the $2\times 2$ identity matrix.  Here, $\lambda$ defines  the 
time-dependent parameter that is to be quenched adiabatically.  
The column vector  $\psi(\vec k)$ defines a two component 
fermionic operator. Such an exactly solvable Hamiltonian (with $\alpha=0$) is known to represent
several one- and two-dimensional integrable quantum spin models as the Ising, the XY 
spin chains \ct{lieb61} and the extended Kitaev model in
two dimensions \ct{kitaev06} when the spin
operators are transformed into spinless fermions via the Jordan-Wigner transformation \ct{lieb61}. In the 
present case, however, the nature of interaction of the spin chains we study, necessitates the consideration of a two
sublattice structure, and hence the fermion operator $\psi(\vec k)= (a_{\vec k},b_{\vec k})$ where $a_{\vec k}$ ($b_{\vec k}$) denote the Jordan-Wigner
Fermions for the mode $\vec k$ describing the spins on even (odd) sublattices. 
The excitation energy  of the Hamiltonian is given by
\be
\epsilon_{\vec k}^{\pm} = \alpha c(\vec k) \pm \sqrt {\lambda^2 + |\Delta|^2}.
\ee
The phase diagram of the model in the $\alpha-\lambda$ plane can be easily  obtained.
The presence of the additional term $\alpha c(\vec k)$ plays a non-trivial role in 
determining the phase diagram of the model by making excitation energy zero for 
specific values of the parameter and the wavevector $\vec k$.
For example, if $\Delta({\vec k})=0$ for the wavevector $\vec k_0$, 
we obtain a critical line ($\epsilon_{\vec k}^{-}=0$) given by $\lambda= \alpha c(\vec k_0)$. 

Our interest is in  the defect generation when the parameter $\lambda$ is quenched in a linear fashion  as $t/\tau$ from
$ t \to -\infty$ to $t \to +\infty$ and the system is swept across the critical line.
Let us assume that at  $ t \to -\infty$, the system is prepared in its ground state $|1_{\vec k} \ket$  such that
$a_{\vec k}^{\dagger}a_{\vec k} |1_{\vec k} \ket =1$ and  $b_{\vec k}^{\dagger}b_{\vec k}|1_{\vec k} \ket =0$ for any $\vec k$. 
For an adiabatic dynamics, the  expected final
state is $|2_{\vec k}\ket$ defined as $b_{\vec k}^{\dagger}b_{\vec k} |2_{\vec k} \ket =1$.  
In course of dynamics, a general state describing the system at an
instant $t$ can be put in the form $\psi_{\vec k} (t) = C_{1,\vec k}(t) |1_{\vec k}\ket + C_{2,\vec k}(t) |2_{\vec k}\ket$ 
where the time dependent
coefficients satisfy the  Schrodinger equation
\ba
i  \frac{\partial}{\partial t} \left( \begin{array}{c} C_{1,\vec k} \\C_{2,\vec k}
\end{array} \right)
&=& H_{\vec k}(t)\left( \begin{array}{c} C_{1,\vec k} \\
C_{2,\vec k}
\end{array} \right) \nonumber \\
&=&\left(
\begin{array}{cc} \lambda(t)& \Delta(\vec k) \\ \Delta(\vec k)^*&-\lambda(t) \end{array} \right)
\left( \begin{array}{c} C_{1,\vec k} \\ C_{2,\vec k}
\end{array} \right)
\ea
with $\hbar$ set equal to unity. Also, the time dependent
parameter $\lambda(t)$ varies as $\lambda_0 t/\tau$ and the initial conditions are $|C_{1,\vec k}(t\to-\infty)|^2 =1$ and
$|C_{2,\vec k}(t\to-\infty)|^2 =0$. 
 It is worth noting that although the term $\alpha c(\vec k) \hat {1}$ plays a crucial role
in determining the critical line or the gapless regions, it does not show up
in the time-dependent Hamiltonian $H_{\vec k}(t)$ which dictates the temporal evolution.
This is because  the identity operator commutes with all the other terms of the Hamiltonian at
every instant of time.  The term $\alpha c(\vec k) \hat {1}$ influences the dynamics only up to a phase factor and 
is hence truly irrelevant in deciding the non-adiabatic behavior. In this sense, the time dependent Hamiltonian 
$H_{\vec k}(t)$ does not capture the passage through quantum critical lines and gapless phases generated 
by $\alpha c(\vec k)$! The Schrodinger equations (3) describing the dynamics of the system effectively 
boils down to  the standard Landau-Zener problem (LZ) \ct{landau} of two time-dependent levels 
$ \pm \sqrt {\lambda(t)^2 + |\Delta|^2}$ (not the levels given in Eq.~2) approaching each other in a linear 
fashion with a minimum gap $2 |\Delta|$ at time $t=0$.
The probability of excitation in the final state is given by Landau Zener formula \ct{landau,sei} 
$p_{\vec k} =  |C_{1,\vec k}(t\to \infty)|^2 = \exp(- (\pi \Delta(\vec k)^2  \tau )/\lambda_0)$ .

Let us assume that the parameter $\Delta(\vec k)$ vanishes at a quantum critical point for $\alpha=0$ as 
$\Delta (\vec k) \sim |\vec k -\vec k_0|^{z_2}$ where $\vec k_0$ is the critical wave vector.  
Noting that in the adiabatic limit of large $\tau$, only modes close to
$\vec k_0$ contribute, the defect density in the final state is given by 
\ba
 n &=& \frac 1  {(2\pi)^d} \int_{BZ} p_k ~d^dk \nonumber\\
 &=&\frac 1  {(2\pi)^d} \int_{BZ} d^dk~ \exp(- (\pi |\vec k -\vec k_0|^{2 z_2} \tau)  /\lambda_0)\nonumber\\
&\sim& \frac 1 {\tau^{d/2z_2}}
\ea
The scaling of the density of defects  hence depends only on the exponent $z_2$ as observed 
previously in references \ct{divakaran09} in the context of quenching through a multicritical point.
 The situation where the parameter $\Delta (\vec k)$ does never vanish during the temporal evolution
of $H_{\vec k}(t)$ is far more interesting. If the parameter $\Delta$ attains
the  minimum value $\Delta_0$ for some wave vector $k_0$ satisfying the scaling form 
$\Delta^2 = \Delta_0^2 + \delta|\vec k - \vec k_0|^{2 z_2}$,  the nonadiabatic transition probability  will show
an exponential behavior and the defect density will scale as
$ n \sim \exp (- \pi \Delta_0^2 \tau/\lambda_0)/\tau^{d/2z_2}$. 
As discussed already, the scaling of the defect density satisfy the same scaling form for all values of $\alpha$.   
This exponential decay of defect even in passage through a quantum critical point is the key  feature of this communication.

Question is whether it is possible to find a Hamiltonian which gets mapped to the Eq.~(1).
To show this we consider a spin-1/2 quantum XY spin with a two sublattice
structure in the
presence of a three-spin interaction and a staggered magnetic field $h$ given by the Hamiltonian 
\ba
&H& = -h\sum_i (\sigma_{i,1}^z -\sigma_{i,2}^z) - J_1 \sum_i (\sigma^x_{i,1}\sigma_{i,2}^x + \sigma^y_{i,1}\sigma_{i,2}^y) \nonumber\\
&-&J_2 \sum_i (\sigma^x_{i,2}\sigma_{i+1,1}^x + \sigma^y_{i,2}\sigma_{i+1,1}^y) 
- J_3 \sum_i (\sigma^x_{i,1} \sigma_{i,2}^z \sigma^x_{i+1,1} \nonumber\\
&+&\sigma^y_{i,1} \sigma_{i,2}^z \sigma^y_{i+1,1}) -J_3 \sum_i (\sigma_{i,2}^x \sigma_{i+1,1}^z \sigma_{i+1,2}^x \nonumber\\
&+& \sigma_{i,2}^y \sigma_{i+1,1}^z \sigma_{i+1,2}^y),
\label{ham2}
\ea
where  $i$ is the site index and the additional subscript $1 (2)$ defines the odd (even) sublattice. The
parameter $J_1$  describes the XY interaction between spins on sublattice $1$ and $2$
while  $J_2 $ describes the XY interaction between spins on  sublattice  $2$ and $1$ 
such that  $J_1$ is not necessarily equal to $J_2$. The parameter  $J_3$, chosen to be
positive throughout, denotes the three spin interaction. Some variants of  the
Hamiltonian (\ref{ham2}) were studied previously \ct{zvyagin06}.
This spin chain is exactly solvable in terms of Jordan-Wigner fermions \ct{zvyagin06}
defined on even and odd 
sublattices as $
\sigma_{i,1}^{+}=\left[\prod_{j<i}(-\sigma_{j,1}^{z})(-\sigma_{j,2}^{z})\right]a_{i}^{\dagger},$
and $
\sigma_{i,2}^{+}=\left[\prod_{j<i}(-\sigma_{j,1}^{z})(-\sigma_{j,2}^{z})(-\sigma_{i,1}^{z})\right]b_i^{\dagger},
$
where 
$\sigma_{i,1}^{z}=2a_{i}^{\dagger}a_{i}-1$  and $\sigma_{i,2}^{z}=2b_{i}^{\dagger}b_{i}-1$.
 The Fermion operators $a_{i}$ and $b_{i}$ can be shown to satisfy Fermionic anticommutation relations.

In terms of the Jordan-Wigner Fermions, the Hamiltonian (5) can be recast in the Fourier space to the form
given in Hamiltonian (1) with

\ba 
H_k= \alpha  \cos k \hat 1 - \frac 1 {2}\left[ \begin{array}{cc} \lambda & -(1+ \gamma e^{-ik}) \\
-(1+ \gamma e^{+ik}) & -\lambda \end{array} \right],  \non \\
& & \label{kit_mat} \ea
where $\psi_k^{\dagger} = (a_k^{\dagger}, b_k^{\dagger})$ and we have set $\lambda = h/J_1$, $\alpha = J_3/J_1$ and $\gamma= J_2/J_1$.  The excitation energy is  obtained
as 
\be
\epsilon_{k}^{\pm} = \alpha \cos k  \pm \frac 1 {2} \sqrt{\lambda^2 + 1 + \gamma^2 + 2 \gamma \cos k}
\ee
Comparing  with the  Hamiltonian (1), we also identify
$c(k) = \cos k$ and $|\Delta(k)|^2 = 1 + \gamma^2 + 2 \gamma \cos k$. The phase diagram of the model for
both $\gamma=1$ and $\gamma \neq 1$ are shown 
in Fig.~1.  For $\gamma \neq 1$, excitation energy $\epsilon_k^+$ vanishes for
the mode $k=\pi$ at the  phase boundary between an antiferromagnetic phase (AF) 
and a gapless phase (GPI) given by $2 \alpha  = \sqrt {\lambda^2 +(1-\gamma)^2}$. 
In the GPI phase, it is always possible to
find a wave vector $k$ for which  $\epsilon_k^+$ vanishes. Similarly, $\epsilon_k^-$ vanishes for the mode
$k=0$ at the phase boundary given by $2 \alpha  = \sqrt {\lambda^2 +(1+\gamma)^2}$ which marks the boundary
between the GPI and the second gapless phase (GPII). In GPII, both $\epsilon_k^+$ and $\epsilon_k^-$ vanish
for some wave vector, so that there are four Fermi points  in contrast to two Fermi points in GPI. 
The transition between GPI and GPII phases is a special quantum phase transition that  involves 
doubling the number of Fermi points \ct{qptfermi}. For $\gamma =1$,
on the other hand, we arrive at a simplified form $|\Delta(k)|^2 = 2 + 2  \cos k= 4 \cos^2 (k/2)$. 
The phase boundaries between the antiferromagnetic phase and GPI phase, and the GPI and GPII 
phase are given by $2\alpha = \lambda$ and $2 \alpha = \sqrt {\lambda^2 + 4}$, respectively. 
It is noteworthy that for the case $\gamma=1$, the parameter $|\Delta|^2$
vanishes at the AF-GPI phase boundary for $k=\pi$ and any $\alpha$  whereas in the anisotropic case,
never does it vanish at the quantum transitions!

\begin{figure}
\ig[height=1.2in]{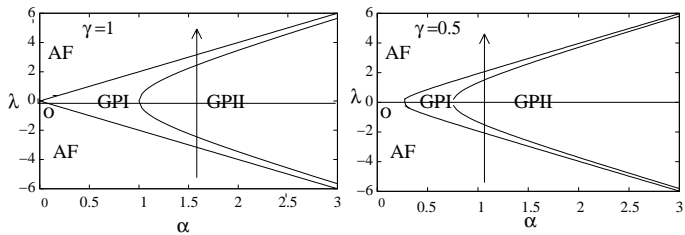} 
\caption{Phase diagram of the Hamiltonian (5) in the $\alpha-\lambda$ plane for  $\gamma =1$ and $\gamma=0.5$. Different phases
are discussed in the main text. The vertical line shows the direction of quenching.}
\label{differenttau}
\textbf{}\end{figure}

In the quenching scheme we employ here, the scaled staggered magnetic field  $\lambda$ is quenched as   
$\lambda_0 t/\tau$ from  $t \to -\infty$ to $+\infty$ with $\alpha \neq 0$ so that the
system is swept across  the quantum critical lines and the gapless phases. Let us set $\lambda_0=1$ for
simplicity. As discussed already, so far as the  dynamics is concerned, the term $\alpha \cos k$ is 
irrelevant and  the Schrodinger equation essentially reduces to a two level LZ problem. 
Note that in both the limits $\lambda  \to \pm \infty$, the spins
should be in a perfect antiferromagnetic orientation in the $z$-direction  and wrongly oriented spins 
in the final state at $t \to \infty$ ($\lambda \to \infty$) are the defects. 

Using the LZ transition formula we find that the probability of excitation in the final state is given by $p_k = \exp (- \pi |\Delta(k)|^2 \tau)
=\exp(-4\pi\cos^2 (k/2)\tau)$ for  $\gamma=1$. In the adiabatic limit of $\tau \to \infty$, only the modes close to 
$k = \pi$ (for which $|\Delta(k)|^2$ vanishes) contribute,
so that $p_k$ takes the simplified from  $p_k = \exp(-\pi (\pi-k)^2 \tau)$, and hence the defect density in the final state is
given by
\be
n = \frac 1 {\pi} \int_0^{\pi} dk \exp(-\pi (\pi-k)^2 \tau) \sim \frac 1{\sqrt \tau}.
\ee
For $\gamma \neq 1$, on the other hand, $\Delta$ is minimum for $k =\pi$ on the AF-GPI phase
boundary given by $|\Delta_0|^2 = |1-\gamma|^2$ so that the defect density

\ba
n &=& \frac 1 {\pi}\int_0^{\pi} dk \exp(-\pi |1+ \gamma^2 + 2 \gamma \cos k| \tau) \nonumber\\
&\sim& \frac{e^{-\pi |\Delta_0|^2\tau}}{\sqrt \tau}.
\ea
We therefore, come across a special situation where the defect density decays exponentially with the rate $\tau$ even though 
the system is swept across the critical lines and gapless phases as long as $\Delta_0 \neq 0$, $i.e.,$ $\gamma \neq 1$.  
We recover the power law scaling in Eq~(8) for $\gamma=1$.
As mentioned already, the scaling behavior shown in Eqs. (8) and (9) are same for
all values of the scaled three spin interaction $\alpha$.  The defect density obtained
by numerical integration of the Schrodinger equations (3) is shown in Fig 2.

\begin{figure}
\ig[height=1.2in,angle=0]{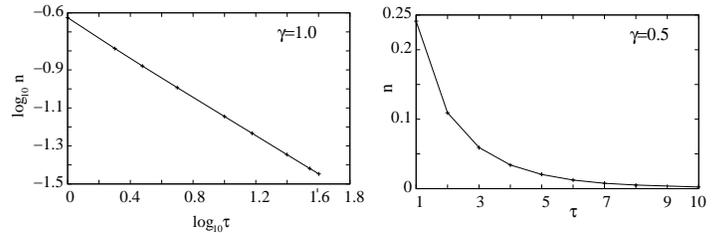} 
\caption{Numerical integration result for the defect density as a function of $\tau$ for $\gamma=1$ and
$\gamma=0.5$. The parameter $\alpha$ is chosen such that the system crosses both the gapless phases 
in the two cases. For $\gamma=1$, we have 
used log-scale and  slope of the straight line is 1/2 indicating a power-law decay of the defect density. 
For $\gamma=0.5$, we have used linear scale to accentuate the exponential fall of the
defect density.}
\end{figure}

It is to be noted that the time evolution governed by $H_k(t)$ is completely 
insensitive to the phase transitions  generated by $\alpha c(k)$. This fact can be understood also using
the following argument: the total number of Fermions for a mode $k$ is a constant of motion as far as the Landau-Zener 
dynamics is concerned. This is because the number operator $n_k = a_k^\dagger a_k + b_k^\dagger b_k$ 
commutes with $H_k(t)$ for all $k$. 
In the initial (final) state
$\lambda \to- \infty$ ( $+\infty$), the expectation value of $n_k$ is  unity which as per the above 
argument is conserved throughout the dynamics. On the other hand, the three spin term gives rise
to the gapless phases where  both the energies $\epsilon_k^+$ and $\epsilon_k^-$ can become negative or
both can become positive for some values of $k$; the true ground state allows $n_k = 2$ or $0$ in those cases. 
However, the instantaneous eigenstates of the time-dependent Hamiltonian  will continue to have $n_k = 1$ 
as explained above and does not reach the true ground state.

Question  remains whether the result presented here should persist in the case of a general 
interacting system.  In ref.\ct{polkovnikov05}, it was shown that the proof of the Kibble-Zurek 
scaling form does  not require the system to be broken 
up into a product of two-level systems which can then be analyzed by
Landau-Zener tunneling formula. The argument only uses translational invariance
and some general scaling arguments, namely, the momentum dependence of the energies and  the 
parameter dependence of the wave functions of those states.
 Extending the argument of ref.\ct{polkovnikov05}  in the present case
demands that the  Hamiltonian should be decoupled into two parts; one time independent $H_1 (\alpha)$ 
and  the other time dependent $H_2 (\lambda(t))$, where $\lambda =t/\tau$ as defined before, and 
 $H_1$  commutes with all the terms of $H_2(\lambda)$. 
It can be shown that $H_1 (\alpha) $ does not influence the dynamics except
for a phase factor and  the scaling form of the density of defects is then given by  scaling form of the
$\lambda$-dependent part of the excitation energy $\delta \omega_{2,k}(\lambda)$ 
 whereas vanishing
of the total excitation energy $\delta \omega_k = \delta \omega_{1,k} (\alpha)+\delta \omega_{2,k}(\lambda)$
leads to the complex phase diagram with critical lines and gapless phases. 
If $\delta \omega_{2,k}(\lambda)$ does not vanish, one
expects an exponential decay of the defect density. 
On the other hand, if $\delta \omega_{2,k}$ vanishes for any mode $k$, 
then the density of defects
decays as a powerlaw with the quenching rate $\tau$. In the present work, we have used an example
of an exactly solved system which satisfies the above conditions. Though possible in principle , 
we believe that finding the example of such an interacting Hamiltonian which satisfies the above 
conditions and show an exponential decay of the defect density is a difficult and open problem.  
Our interest in this work is only to point out a special situation  where one can find exponential 
decay of defect density even during passage through critical regions and we have illustrated the 
possibility using an exactly solvable spin chain with complicated interactions.

{\it Acknowledgement} We acknowledge Diptiman Sen  and G.E. Santoro for very interesting 
and critical comments and Victor Mukherjee for carefully reading the manuscript. 
D.C. acknowledges KVPY, Department of
Science and Technology (Government of India) for support and also G.
Refael for interesting discussions.
AD and UD acknowledge Abdus Salam ICTP, Trieste, Italy, where some part of this work was done.

\end{document}